\documentclass[english]{article}
\usepackage[T1]{fontenc}
\usepackage[latin9]{inputenc}
\usepackage{color}
\usepackage{babel}
\usepackage{amsmath}
\usepackage{amssymb}
\usepackage{graphicx}
\usepackage[colorlinks=true]{hyperref}
\makeatletter
\@ifundefined{date}{}{\date{}}
\newcommand{\ee}{e^+e^-}
\usepackage[font={small,it}]{caption}

\@ifundefined{showcaptionsetup}{}{%
 \PassOptionsToPackage{caption=false}{subfig}}
\usepackage{subfig}
\makeatother

\begin{document}

\begin{titlepage}
\clearpage
\thispagestyle{empty}
\hfill RM3-TH/19-2
\vfill

\begin{center}
\LARGE{Beyond the Standard Model physics at CLIC}
\vspace{1.5cm}

{\large{}Roberto Franceschini}\\
\textit{\small{}Università degli Studi Roma Tre and INFN Roma Tre,
Via della Vasca Navale 84, I-00146 Roma, ITALY}{\small\par}
\par\end{center}

\vspace{1cm}
\begin{abstract}
A summary of the recent results from CERN Yellow Report on the CLIC
potential for new physics is presented, with emphasis on the direct
search for new physics scenarios motivated by the open issues of the
Standard Model.
\end{abstract}

\vfill
{\it Talk presented at the International Workshop on Future Linear Colliders (LCWS2018), Arlington, Texas, 22-26 October 2018. C18-10-22.}
\end{titlepage}
\pagebreak{}

\section{Introduction}

The Compact Linear Collider (CLIC) \cite{CERN-2018-005-M,Robson:2018aa,Roloff:2018aa,1812.02093}
is a proposed future linear $\ee$ collider based on a novel two-beam
accelerator scheme \cite{Aicheler:1500095}, which in recent years
has reached several milestones and established the feasibility of
accelerating structures necessary for a new large scale accelerator
facility (see e.g. \cite{Ruth:747483}). The project is foreseen to
be carried out in stages which aim at precision studies of Standard
Model particles such as the Higgs boson and the top quark and allow
the exploration of new physics at the high energy frontier. The detailed
staging of the project is presented in Ref.~\cite{CLIC-staging-2018-CLICdp-Note-2018-002,Aicheler:2652600},
where plans for the target luminosities at each energy are outlined.
These targets can be adjusted easily in case of discoveries at the
Large Hadron Collider or at earlier CLIC stages. In fact the collision
energy, up to 3 TeV, can be set by a suitable choice of the length
of the accelerator and the duration of the data taking can also be
adjusted to follow hints that the LHC may provide in the years to
come. At present we consider a scheme that is a balanced choice aimed
at fully exploiting the physics potential of the project in a manageable
timeframe. CLIC is foreseen to deliver integrated luminosities
\begin{align*}
1\text{ ab}^{-1} & \text{ at }\sqrt{s}=380\text{ GeV}\\
2.5\text{ ab}^{-1} & \text{ at }\sqrt{s}=1.5\text{ TeV}\\
5\text{ ab}^{-1} & \text{ at }\sqrt{s}=3\text{ TeV}
\end{align*}
which will be collected in an overall period of 27 years, preceded
by a 7 years lead time to construct and commission the first stage
of the machine \cite{CERN-2018-005-M}.

The large amount of data collected at the 380 GeV stage will allow
a substantial improvement of our understanding of the top quark \cite{Abramowicz:2018aa}
and the Higgs boson \cite{Abramowicz:2016zbo} compared to what the
HL-LHC will be able to provide~\cite{Cid-Vidal:2018ab,Azzi:2019aa,Cepeda:2019aa}.
The later stages in the trans-TeV center-of-mass energy regime will
allow to thoroughly explore new physics at the TeV scale, benefiting
from the clean $e^{+}e^{-}$ collision environment. In these stages
further improvements on Higgs physics and other SM properties are
expected, thanks to the large number of Higgs bosons, exceeding one
million, that can be detected from vector boson fusion production
and the enhanced sensitivity that such high-energy collisions will
enjoy to reveal new contact interactions which show up in subtle effects
in reactions involving SM states and can be originated by heavy new
physics.

In the following we give a detailed account of recent work \cite{1812.02093}
done for the direct search of these heavy new physics states. In many
examples we will see how a trans-TeV leptonic collider can significantly
extend the reach of the HL-LHC. These results clearly show how such
high energy leptonic colliders can be \emph{discovery machines}, as
well as delivering a clean environment in which to carry out very
precise measurements useful to test accurate SM theoretical predictions,
as done at previous $\ee$ machines like LEP and SLC.

\section{Open issues in the Standard Model and Direct searches of New Physics}

Several issues remain open in particle physics. Questions can be asked
on the origin of the many parameters of the Standard Model and we
are not able to provide satisfactory answers for the dynamics that
generate them. Some of these parameters may be ``just so'', and
the Standard Model would be a consistent and predictive theory of
fundamental interaction. Even if one takes this view, there are observed
phenomena that are simply not accountable in the Standard Model, such
as the astronomical observations that lead to hypothesize the presence
of a new form of matter, the so-called Dark Matter of the Universe,
or some deep revision of the inner workings of gravitational interactions
on the largest length-scales. Either way we should expect the Standard
Model to be replaced by a more fundamental theory capable of accounting
for the phenomenology of Dark Matter. In the well established theory
of the expanding Universe it appears that the amount of matter in
excess of that of anti-matter was miraculously offset in the initial
conditions of our cosmology to result in the amount of matter that
we observe today. Such improbable coincidence in the history of the
Universe can be made a consequence of dynamics of fundamental interactions,
but requires to extend the Standard Model to find sources of baryon
number violation and additional sources of CP violation. Other parameters
of the Standard Model may even break the longstanding paradigm for
which long-distance observables are not affected by tiny modifications
of microscopic degrees of freedom \cite{Hooft:1979bh,Zeldovich:1967gd}.
Establishing this sensitivity to microscopic physics or finding the
mechanism that protects the Standard Model parameters from being so
sensitive would settle a deep issue in the characterization of fundamental
interactions. Particle colliders have a chance to shed some light
on all these issues, but it should be reminded that they are not the
sole type of experiments \cite{Beacham:2019aa} capable of testing
ideas to address the open issues of the Standard Model. Still they
appear to be the only way forward if we want to explore the high energy
frontier, at which some of the explanations of the above mentioned
phenomena may be discovered. In this respect the possibility afforded
by CLIC to collide point-like particles such as electrons and positrons
accelerated at multi-TeV energies appears particularly interesting
to \emph{directly} test new physics scenarios motivated by the need
to address the open issues of the Standard Model.

\paragraph{Direct discoveries of new particles}

CLIC can probe TeV scale electroweak charged particles well above
the HL-LHC reach. Such new particles are naturally expected because
many of the issues that need to be addressed in the Standard Model
are inherent to the electroweak sector of the theory. For example
particle dark matter candidates can hardly carry any other SM charge
than electroweak. Furthermore the electroweak sector may be able to
accommodate the violation of baryon number, C and CP necessary for
the generation of a net baryon number of the Universe as well as provide
a phase transition and the necessary boundaries between phases at
which to generate the net baryon number. All the problems about the
origin of the masses and mixings of neutrinos and of the other fermions
of the SM are related to the weak interactions. In addition, the Naturalness
Problem is genuinely a question about the peculiarity of weak interactions.
A complete exploration of TeV scale electroweak particles is thus
a priority for particle physics. 

Any such new particle can be produced at CLIC with sizable rate up
to the kinematic limit of 1.5 TeV, and in some cases up to 3 TeV via
single production mechanisms. Depending on the decay channels, different
detection strategies are possible. 

When new particles decay into standard final states featuring prompt
jets, leptons and photons they give rise to signatures which can be
relatively easily be distinguished from backgrounds.  Indeed, background
from SM processes usually have cross-section comparable with that
of the signal, as they are produced via the same electroweak interactions.
This is the key advantage of lepton colliders over hadron colliders
that makes CLIC out-perform the HL-LHC. In most cases the signals
can be isolated so clearly that it is possible to measure the properties
of new particles, such as mass and spin, and even test concrete models
of physics beyond the Standard Model by checking some of their key
predictions on new physics particles properties. Examples of reach
for direct discoveries are discussed in detail in Ref.~\cite{1812.02093},
e.g. for new scalars and for several examples from supersymmetric
models. We highlight Section 4.4 of Ref.~\cite{1812.02093} which
contains studies, not summarized here for brevity of this contribution,
on the test of models which predict the Higgs boson mass from other
model parameters, such as the MSSM, and tests of other predictions
from models of Neutral Naturalness such as Twin Higgs models \cite{Chacko:2017aa}.
Other examples of direct reach for discovery for models of particle
Dark Matter, baryogenesis and neutrino mass generations are discussed
in detail in the following. We should recall that the CLIC potential
to directly explore new physics is extensively documented in the literature,
and in particular in early Refs.~\cite{Linssen:2012ss,CLIC-Physics-Working-Group:-E.Accomando:2004aa},
often taking supersymmetric particles as benchmarks. These were studies
from before the Run2 of the LHC, hence cannot profit from the vantage
point we reached today in the exploration of the TeV scale. In the
following we will try to use as much as possible simplified models
that extend the SM in the spirit of well known BSM models, but do
not necessarily carry the whole bag of phenomenological consequences.
For instance we will discuss WIMP-like Dark Matter searches by just
adding a new electroweak state to the SM matter content, or we will
study generic singlet scalar particles without attaching them to any
particular model of the large class in which they may arise (e.g.
in models of electroweak symmetry breaking or baryogenesis scenarios).

When the new particles give rise to non-standard signatures, e.g.
because they decay at a macroscopic distance in the detector volume,
it is still possible to isolate these signals thanks to the clean
environment typical of $\ee$ colliders. Relevant examples of this
kind of signatures include Higgs boson rare decays to long-lived particles,
Higgsino Dark Matter and the search for WIMP baryogenesis models which
are further discussed below. In this case as well the models we present
are representative models of some large class of BSM scenarios and
the results we present allow recasts of the provided information for
other new physics scenarios.

\paragraph{Extended Higgs Sector}

\begin{figure}
\begin{centering}
\includegraphics[width=0.49\linewidth]{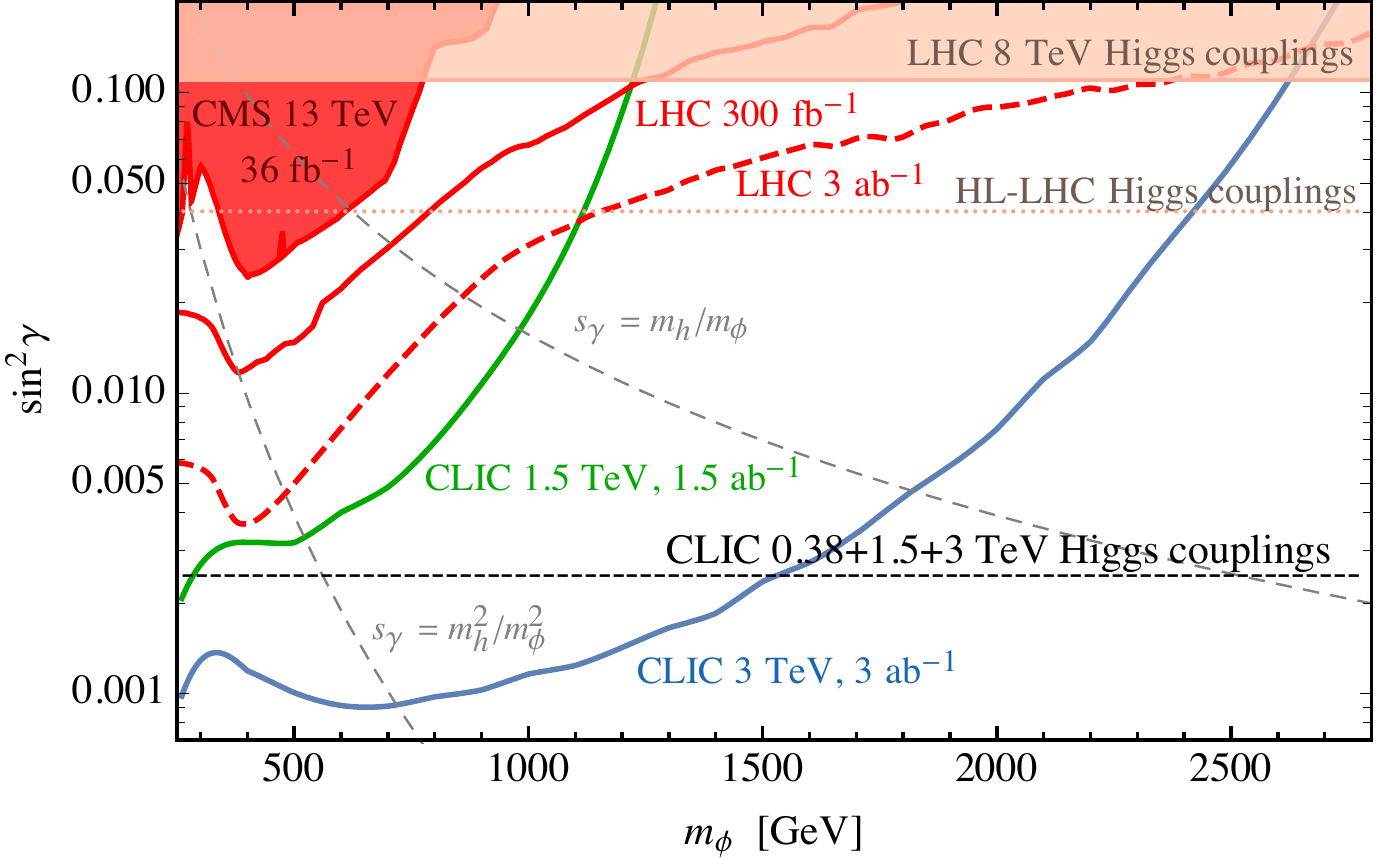}
\par\end{centering}
\caption{\label{fig:Reach-for-Singlet}CLIC reach for new scalar singlets direct
observation (blue and green solid lines)\cite{Buttazzo:2018aa}. Constraints
from measurements of 125 GeV Higgs boson couplings are reported as
horizontal lines. LHC expectations for both direct and indirect sensitivity
are reported as well.}
\end{figure}

Understanding the nature of the Higgs boson is one of the key elements
for a full understanding of the electroweak interactions and in particular
of the breaking of electroweak symmetry. A very important question
is if the Higgs is the unique scalar particle at the weak scale, or
if instead an extended scalar sector exists. For this reason it is
a key target for future colliders to investigate the existence of
additional Higgs bosons at the TeV scale, which may be the first important
step to unravel the mystery of electroweak symmetry breaking and the
origin of the weak scale. 

A prototypical example of extended Higgs sector is the extension of
the SM with a new scalar. A particularly challenging case is the one
in which the new scalar has no gauge interactions and interacts with
SM only through the Higgs boson portal. This kind of scalar is usually
referred as a ``singlet'' scalar and arises in concrete models such
as the Next-to-Minimal Supersymmetric Standard Model (NMSSM), non
minimal Composite Higgs models as well as Twin Higgs models from ``neutral
naturalness'' solution to the hierarchy problem of the weak scale.
In addition, a new such scalar may affect the Higgs boson potential
and alter the nature of the phase transition between broken and unbroken
electroweak symmetry in the early Universe, thus playing a role in
the generation of a net baryon number. 

A concrete study of direct production of new scalar singlet at CLIC
is summarized in Fig.~\ref{fig:Reach-for-Singlet}. CLIC sensitivity
to direct production of a new scalar singlet extends well beyond the
TeV mass scale at which these new singlets are most motivated. For
a mixing between the singlet and the Higgs $\sin^{2}\gamma<0.24\%$
the new singlet has to be heavier than 1.5 TeV. Furthermore if a singlet
of \emph{any mass} has mixing $\sin^{2}\gamma>0.24\%$ it would result
in deviation in the single Higgs couplings to SM gauge bosons and
fermions in excess of 2 standard deviations for the expected accuracy
of Higgs couplings determinations at CLIC. These studies are discussed
in detail in Sec. 4.2 of Ref.~\cite{1812.02093}, where their implications
on concrete models are also worked out. It is found that in the case
of the NMSSM CLIC can exclude a new scalar lighter than 1.5~TeV for
values of $\tan\beta<4$, where the NMSSM is most motivated. For Twin
Higgs models direct search bounds generically rule out new scalars
below 2 TeV for values of the dynamical scale of the model $f<2\,\text{TeV}$
where the model is most motivated. Furthermore, the study of Higgs
boson couplings excludes $f<4.5\text{ TeV}$ if one assumes that the
mass of the scalar is equal or greater than $f$, as expected for
a composite scalar. Similarly to what happens for the NMSSM bound,
such a constraint on the Higgs compositeness scale $f$ would push
the model out of its most motivated region of parameters space. We
point out to the reader that further studies on extended Higgs sectors
can be found in Sec.~4.3 of Ref.~\cite{1812.02093}, which reports
results from Ref.~\cite{Azevedo:2018aa} on several concrete models
featuring extended scalar sectors with multiple doublets and singlets
fields, e.g. the 2-Higgs-Doublets model (2HDM). CLIC can probe the
existence of such new scalars both by direct production of new scalars
and by indirect effects they have on the 125 GeV Higgs boson couplings.
The expected reach extends up to and beyond 1 TeV, improving dramatically
on the HL-LHC searches. Also for light singlet scalars, studied in
Ref.~\cite{Frugiuele:2018aa} and presented in \cite{1812.02093},
it is found that the HL-LHC reach is significantly extended by each
stage of CLIC. In particular these studies for light scalar searches
highlight the importance of scalar-strahlung searches at the first
stage of CLIC (and/or lower energies lepton colliders \cite{Drechsel:2018aa,Wang:2019aa})
as well as the direct searches for specific final states in the 125
GeV Higgs boson decays, such as $h\to4b$ \cite{Liu:2016lq}.  All
in all CLIC is able to thoroughly test extended Higgs sectors and
rule out new scalars up to multi-TeV masses. Both direct and indirect
signatures can be successfully pursued yielding stringent bounds on
new scalar particles that improve by almost one order of magnitude
in mass scale compared to the HL-LHC.

\paragraph{Composite Higgs}

\begin{figure}
\begin{centering}
\subfloat[\label{fig:ch}$5\sigma$ discovery contours for Higgs compositeness
in the $(m_{*},g_{*})$ plane, overlaid with the $2\sigma$ projected
exclusions from HL-LHC. Shaded areas are labeled as ``pessimistic''
and ``optimistic'' depending if operators coefficients were taken
a factor 2 larger or smaller than the base estimates from the $(m_{*},g_{*})$
value.]{\centering{}\includegraphics[width=0.45\linewidth]{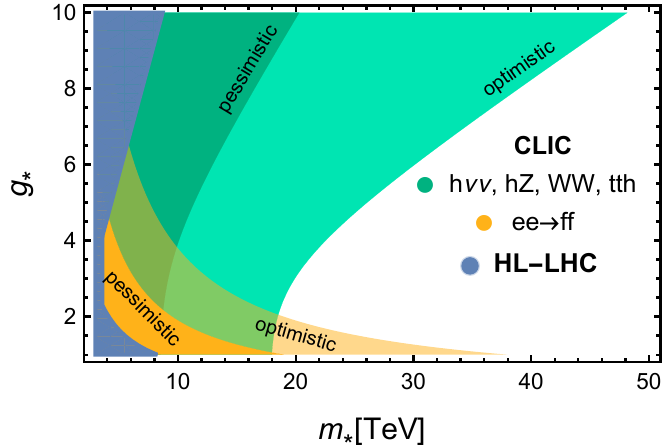}
}~~\subfloat[\label{fig:ch-1}The $5\sigma$ top quark compositeness discovery
contours in the $(m_{*},g_{*})$ planes from studies of $t\bar{t}$
and $t\bar{t}h$ final states. Darker and lighter shaded areas corresponds
to the variations of the size of the operators coefficients by a multiplicative
factor of 2 or 1/2 on top of the baseline expectation form the values
of $m_{*}$ and $g_{*}$.]{\centering{}\includegraphics[width=0.45\linewidth]{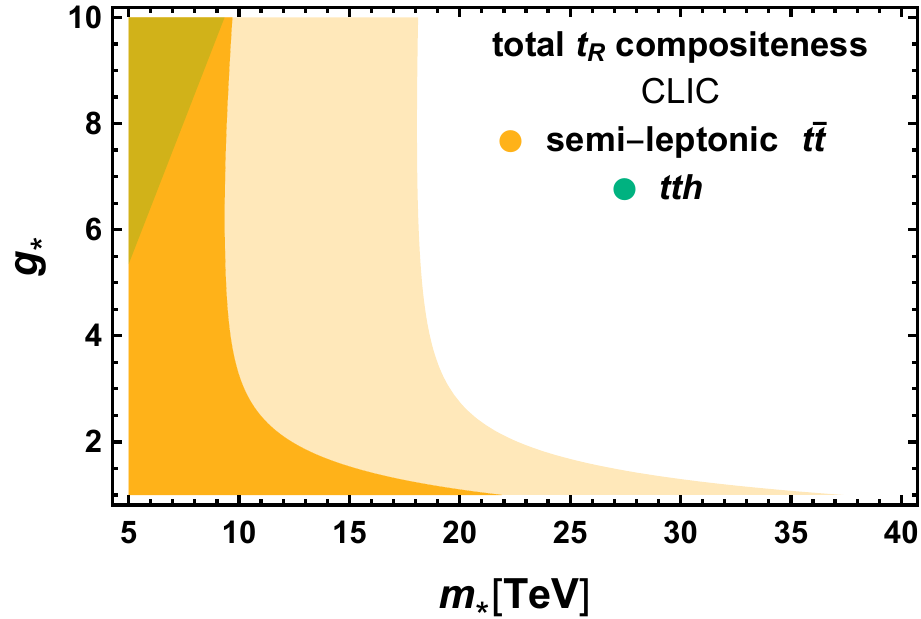}
}
\par\end{centering}
\caption{Composite Higgs reach from Higgs boson, top quark and Drell-Yan studies
taken from Refs.~\cite{1812.02093} and \cite{Abramowicz:2018aa}.}
\end{figure}

The Higgs boson is the only scalar particle that is predicted in the
SM to be exactly point-like. Therefore it is interesting to investigate
if instead it is an extended composite object and, if it is, to determine
its geometric size $l_{H}$. Discovering the composite nature of the
Higgs would be a crucial step towards the understanding of the microscopic
origin of the 
electroweak symmetry breaking phenomenon. Higgs compositeness might
also provide a screening of SM Higgs mass parameter from details of
the ultra high energy modes, as the Standard Model would be replaced
by a whole new fundamental theory of the electroweak sector at distances
shorter than the size of the Higgs boson. At a phenomenological level
a composite Higgs would manifest itself at CLIC through contact interactions,
e.g. the $d=6$ SMEFT operators, suppressed by two powers of the Higgs
compositeness scale $m_{*}\sim1/l_{H}$. The operator coefficients
are enhanced or suppressed, relative to the naive $1/m_{*}^{2}$ scaling,
by positive or negative powers of a parameter ``$g_{*}$'' representing
the coupling strength of the composite sector the Higgs emerges from
\cite{Giudice:2007yg}. These rules provide estimates for the operator
coefficients in the $(m_{*},g_{*})$ plane and allow us to translate
the CLIC sensitivity to the SMEFT into the discovery reach on Higgs
compositeness, as displayed in Figure~\ref{fig:ch} from Ref.~\cite{1812.02093}.
The projected HL-LHC exclusion (as opposite to discovery lines shown
for CLIC) reach is also shown in the figure for unit $c$-coefficients.
The dramatic improvement achieved by CLIC at small and intermediate
$g_{*}$ is due to the high-energy stages that allow for a very precise
determination of the $c_{HW}$, $c_{HB}$, $c_{2W}$ and $c_{2B}$
SMEFT Wilson coefficients. Single Higgs boson couplings measurements
are instead provide the most stringent constraints at large $g_{*}$.

This example shows magnificently the complementarity of the searches
for new physics that can be attained at CLIC by exploiting the \emph{precision
and the mass reach} that CLIC will provide. Measurements from high-intensity
studies, in this case studies of copiously produced Higgs bosons,
probe one combination of the two characteristic parameters of this
scenario, while the other combination is probed by less copious events
at high invariant mass in the later stages of CLIC. It should be remarked
that a similar conclusion can be reached in the study of the general
$d=6$ SMEFT for universal theories, in which the study of high energy
processes allows to put constraints on quantities such as the ``S
parameter'' \cite{Barbieri:2004ek,Altarelli_1992,Altarelli_1991,Peskin_1992,Peskin_1990}
that is traditionally measured in high-intensity experiments at the
$Z$ pole~\cite{Franceschini:2017ab,Henning:2018aa}. This complementarity
of approaches is a great advantage of high energy lepton colliders.

Within the compositeness context, and in connection with the Naturalness
Problem, top quark compositeness can also be considered. It produces
SMEFT operators in the top sector that can be probed by measuring
the top Yukawa coupling and, very effectively, by $t{\overline{t}}$
production at high-energy CLIC. The reach in the ``total $t_{R}$
compositeness'' scenario is displayed on Figure~\ref{fig:ch-1}.
For further details, and for a similar result in the case of ``partial
top compositeness'', see Sec.~2.1 of Ref.~\cite{1812.02093} and
Sec.~10.2 of Refs.~\cite{Abramowicz:2018aa}.

\paragraph{Dark Matter}

\begin{figure}
\begin{centering}
\subfloat[\label{fig:reach-DM}95\% excluded masses for new electroweak $n$-plet
states with hypercharge $Y$. The exclusion for each state denoted
by (1,n,$Y$) at CLIC Stage2 and Stage3 is presented in green and
yellow bar \cite{Di-Luzio:2018aa}.]{\begin{centering}
\includegraphics[width=0.49\linewidth]{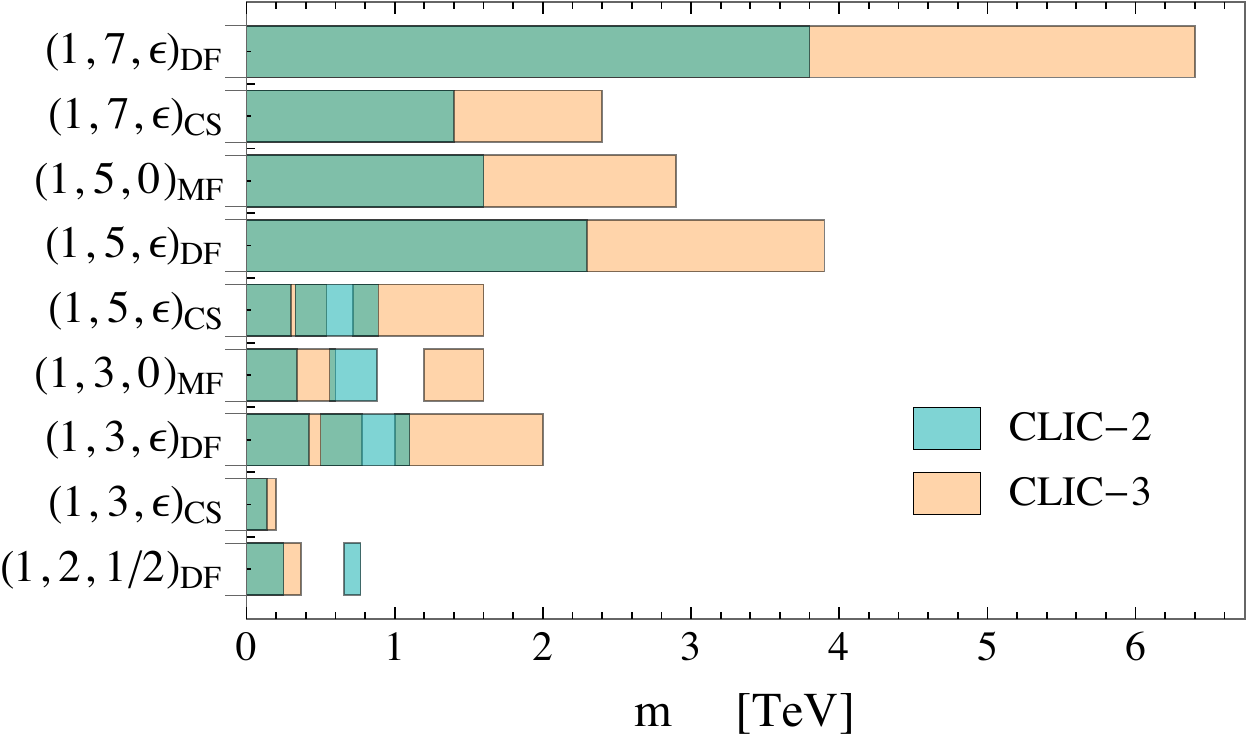}
\par\end{centering}
}~~~\subfloat[\label{fig:reach-higgsino}95\% excluded region for a new physics
signal from a pure Higgsino in the mass-lifetime plane. The black
dashed line denotes the theory prediction for a the lifetime of a
pure Higgsino doublet. Green, Yellow and Blue areas correspond to
3 TeV, 1.5 TeV and 380 GeV CLIC runs expected exclusions. ]{\begin{centering}
\includegraphics[width=0.44\linewidth]{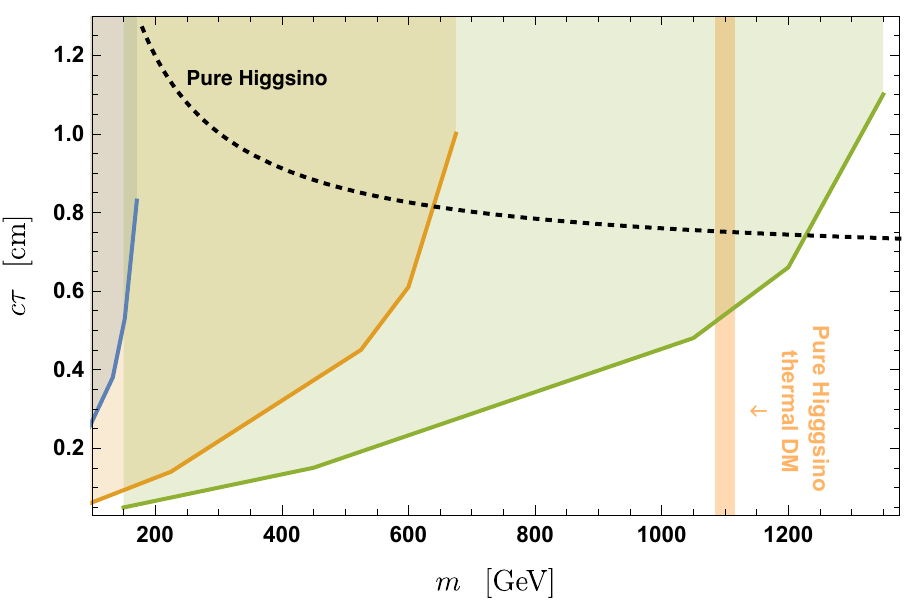}
\par\end{centering}
}
\par\end{centering}
\caption{Reach of direct searches for Dark Matter.}
\end{figure}

The nature of dark matter in the Universe is a great mystery. Very
little is known about the particle properties of dark matter and a
large host of models provide viable dark matter candidates. A particularly
compelling candidate is the so-called Weakly Interacting Massive Particles
(WIMP) that naturally emerge from the standard cosmological history
of the Universe as possible thermal relics that stop being in equilibrium
with the plasma of the early Universe at a temperature roughly one
order of magnitude below their mass and from that moment onward remain
as relics in the Universe, interacting with the SM particles only
through gravity, ultimately shaping the formation of galaxies and
other cosmic structures. Thermal production of WIMPs can yield the
observed abundance of dark matter for masses $M_{\text{WIMP}}\simeq\text{TeV}\left(\frac{g_{\text{SM,DM}}}{g_{\text{weak}}}\right)^{2}$
where $g_{\text{DM,SM}}$ roughly denotes the strength of the couplings
of processes that keep the dark matter in equilibrium, e.g. $DM\,DM\rightleftarrows SM\,SM$,
and $g_{\text{weak}}$ is the coupling strength of the SM weak interactions.

Despite these requirements on the particle nature of the Dark Matter
a large set of possibilities exists even if one restricts to consider
weakly interacting massive particles. In Sec.~5 of Ref.~\cite{1812.02093}
a comprehensive strategy is outlined to test a wide range of possible
situations in which the Standard Model is extended by a WIMP and by
other states possibly members of the same weak interactions multiplet
or as independent state.

The approach we follow to study Dark Matter phenomenology by simply
specifying masses and quantum numbers of new states has been called
``Minimal DM''~\cite{Cirelli:2005uq,Di-Luzio:2015ab,Mitridate:2017fk}.
This approach has shown that new particles that can live also in very
large (up to the $7$-plet) representations of the SM $SU(2)$ group
can give viable Dark Matter candidates. All this variety of weakly
charged states are a target for future colliders. CLIC can probe them
in several ways. First, one can perform model-independent indirect
searches for new EW states by studying their radiative effects on
the EW pair-production of SM particles, obtaining the $95\%$~CL
sensitivities reported in Figure~\ref{fig:reach-DM} taken from results
of Ref.~\cite{Di-Luzio:2018aa}. The sensitivity reaches the thermal
mass (i.e., the one which is needed in order to produce the observed
thermal abundance) in the case of the Dirac fermion triplet candidate
$(1,3,\epsilon)_{{\rm {DF}}}$. Second, one can exploit the fact that
the charged component of the Minimal DM multiplet is long-lived, with
a macroscopic decay length. Its distinctive signature is thus a ``stub''
track, which can be long enough to be seen if the particle is light
enough to be sufficiently boosted. Figure~\ref{fig:reach-higgsino}
shows that CLIC can discover the thermal Higgsino at $1.1$~TeV with
this strategy.

In addition, it should be noted that CLIC is also sensitive to DM
models that fall outside the Minimal DM paradigm, such as co-annihilation
scenarios, in which two almost degenerate states can scatter into
Standard Model states with a much stronger interaction than each of
them singly. Models that exploit the presence of multiple states,
such as the Inert Doublet model \cite{Barbieri:2006dq,Deshpande_1978,Cao:2007aa},
can also be thoroughly explored at CLIC, extending significantly the
domain of the parameters space probed in comparison to the HL-LHC
capabilities \cite{Kalinowski:2018ac,Kalinowski:2018ab}. Details
on these and other models are presented in Ref.~\cite{1812.02093}.
Here we content with stating that in general CLIC can effectively
probe DM models with a sufficient mass-splitting to produce signals
featuring prompt jets, leptons and photons plus missing momentum.

\paragraph{Baryogenesis and electroweak phase transition}

\begin{figure}
\begin{centering}
\subfloat[\label{fig:ewbg-reach}Reach on the Higgs plus singlet model for electroweak
baryogenesis from \cite{No:2018aa}. First order phase transition
is allowed in the parameters space marked by green points. Regions
outside the blue (black) dashed lines are excluded by CLIC direct
search for $S\to HH$ (triple Higgs coupling measurement).]{\begin{centering}
\includegraphics[width=0.38\linewidth]{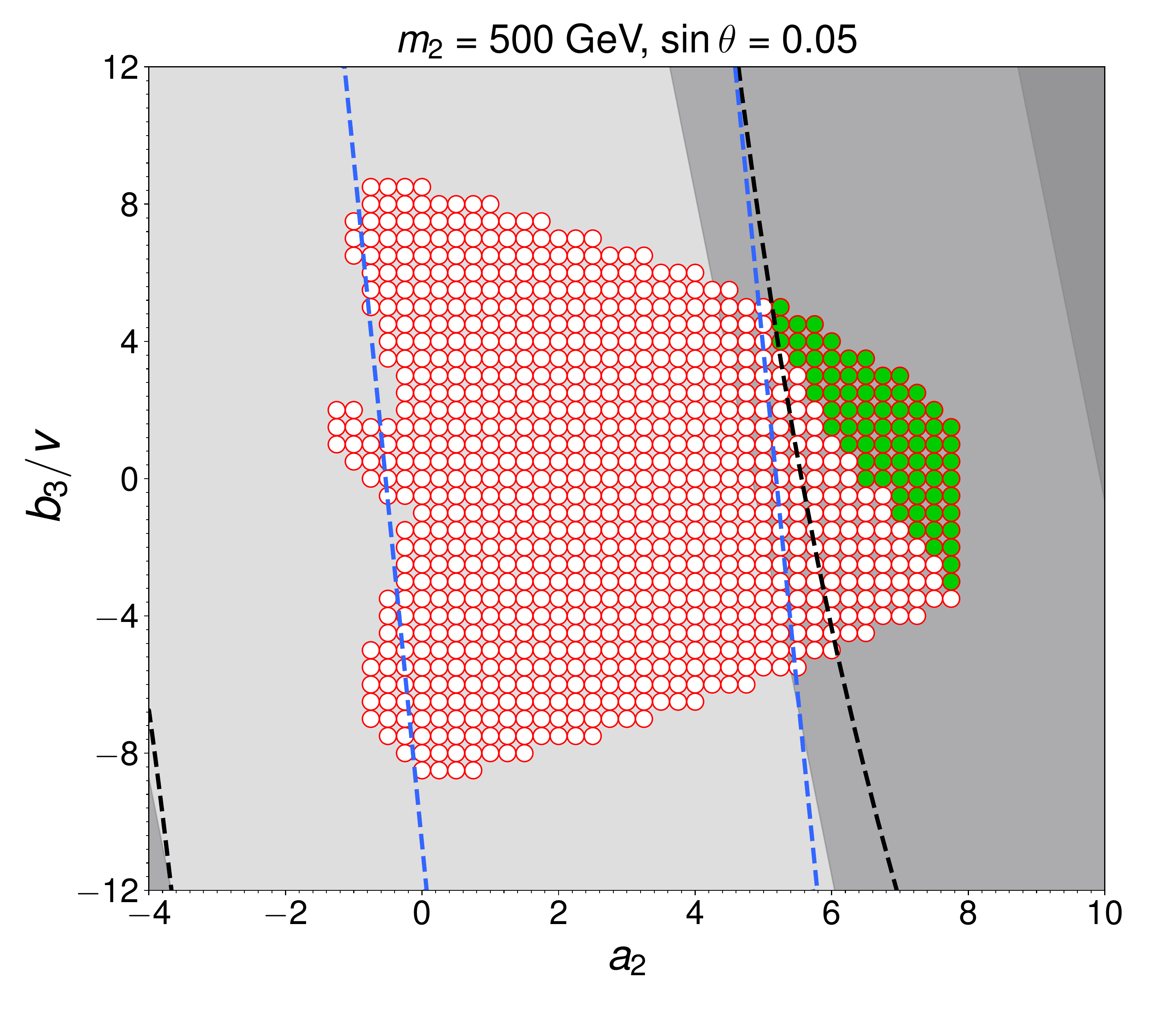}
\par\end{centering}
}~~\subfloat[\label{fig:ewbg-reach-1} Reach on the Higgs plus singlet model for
electroweak baryogenesis from \cite{Buttazzo:2018aa}. First order
phase transition is allowed in the parameters space marked by shaded
regions. Overlaid are iso-lines (dashed) for single Higgs couplings
modification, denoted by $\Delta\kappa$, and triple Higgs couplings
deviations iso-lines, denoted by $\Delta\lambda_{hhh}$. Iso-lines
for the number of produced events $\ee\to SS+X$ are reported as thick
and thin lines.]{\begin{centering}
\includegraphics[width=0.5\linewidth]{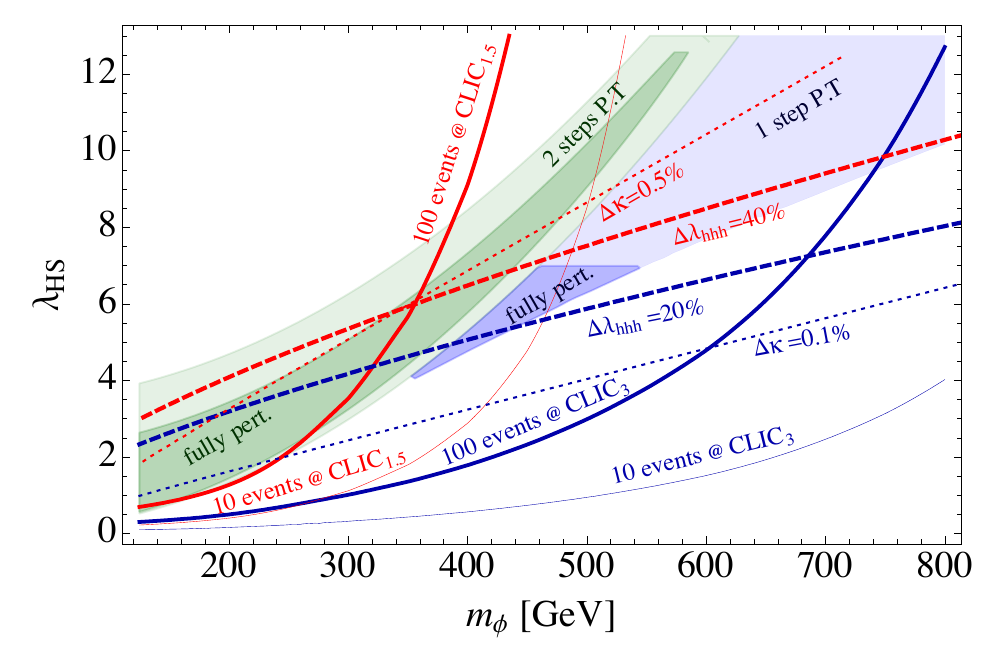}
\par\end{centering}
}
\par\end{centering}
\caption{CLIC reach for the Higgs plus a scalar singlet for electroweak baryogenesis.}
\end{figure}

The mechanism responsible for the origin of the net baryons number
of the Universe is currently unknown, and it might be discovered at
CLIC if it is related with TeV-scale physics. A prominent example
of such dynamical origin of the baryons in the Universe is the ElectroWeak
BaryoGenesis (EWBG) scenario (see \cite{10.1088/978-1-6817-4457-5}
for an introduction). This mechanism requires, among other things,
a considerable modification of the SM thermal Higgs potential, that
should give rise to an EW phase transition of strong first order,
unlike the smooth crossover that is predicted by the SM. This is achieved
through new scalar particles coupled with the Higgs, which can modify
its potential at tree level or via loops. These particles, as they
must have some interaction with the Higgs boson, can be probed at
CLIC by precise measurements of the Higgs trilinear coupling and of
single Higgs couplings to SM states, as well as by direct searches.

The most minimal of such models extends the Standard Model by just
adding a new singlet scalar and allowing the most general scalar potential
for a Higgs doublet and the new singlet scalar. This is a suitable
illustrative benchmark because it contains the minimal amount of new
physics (i.e., a scalar singlet $S$) that is needed to achieve a
strong first-order phase transition. Several measurements can be used
to constrain the model at CLIC, as is illustrated in Figure~\ref{fig:ewbg-reach},
taken from \cite{1812.02093,No:2018aa}. The figure displays a slice
of the parameter space of the model for singlet mass $m_{2}=500$~GeV
and singlet mixing with the SM Higgs boson $\sin\theta=0.05$. The
remaining parameters $a_{2}$ and $b_{3}$ are respectively the $|H|^{2}S^{2}$
quadratic portal coupling and the $S^{3}$ trilinear vertex. The allowed
points in the plane are marked as red circles, those for which the
EW phase transition is strong enough are filled in green. All the
green points can be probed both by the trilinear Higgs coupling measurement
(black dashed) and by single $S$ production decaying to $HH\rightarrow4b$
final state (blue dashed). For this choice of mixing, the entire plane
is probed by single Higgs coupling measurements (gray region). Notice
however that the effectiveness of single-Higgs couplings stems from
the fact that the model at hand predicts sharp correlations between
the modification of several Higgs vertices, but this correlations
might be relaxed in other models. The CLIC capability of performing
multiple competitive probes of the scenario instead allows to draw
robust conclusions. Indeed the measurement of the Higgs trilinear
couplings remains a powerful constraint even when $\sin\theta\to0$,
and it allows to exclude models with $m_{S}>450\text{ GeV}$.

In addition it is possible to search at CLIC for pair production of
the singlet $S$, which is mediated directly by the $a_{2}$ coupling,
hence it is allowed even when $S$ and the Higgs boson doe not mix.
In a scenario completely driven by the $a_{2}$ coupling it is possible
to study loop effects on the Higgs potential and put constraints on
the model using the same measurements mentioned above: single Higgs
couplings and triple Higgs couplings. The limits in this scenario
are reported in Figure~\ref{fig:ewbg-reach-1}, where the $a_{2}$
coupling is renamed $\lambda_{HS}$ following the notation of Ref.~\cite{Buttazzo:2018aa}.
This figures also report the regions of parameter space in which the
$\lambda_{HS}$ coupling is large enough, or the singlet is light
enough, that CLIC can produce ten or one hundred $S$ pairs. The number
of events to be produced to put a bound or discover $S$ depends on
the specific decay of the singlet $S$, but these numbers are reasonable
estimates for an $\ee$ collider and experimental signatures with
moderate level of background. From these considerations CLIC is expected
to be sensitive to pair production of singlet scalars related to EWBG
for a large region of the parameter space in which the phase transition
can be first order.

\begin{figure}
\begin{centering}
\subfloat[\label{fig:Excluded-cross-section-h2PiV}Excluded cross-section time
Higgs branching ratio in the exotic mode $h\to\pi_{v}\pi_{v}$. The
reference rate for CLIC 3 TeV is $\sigma_{h}\sim0.4\text{ pb}$ in
this study \cite{Kucharczyk:2018aa,Kucharczyk:2625054}.]{\includegraphics[width=0.34\linewidth]{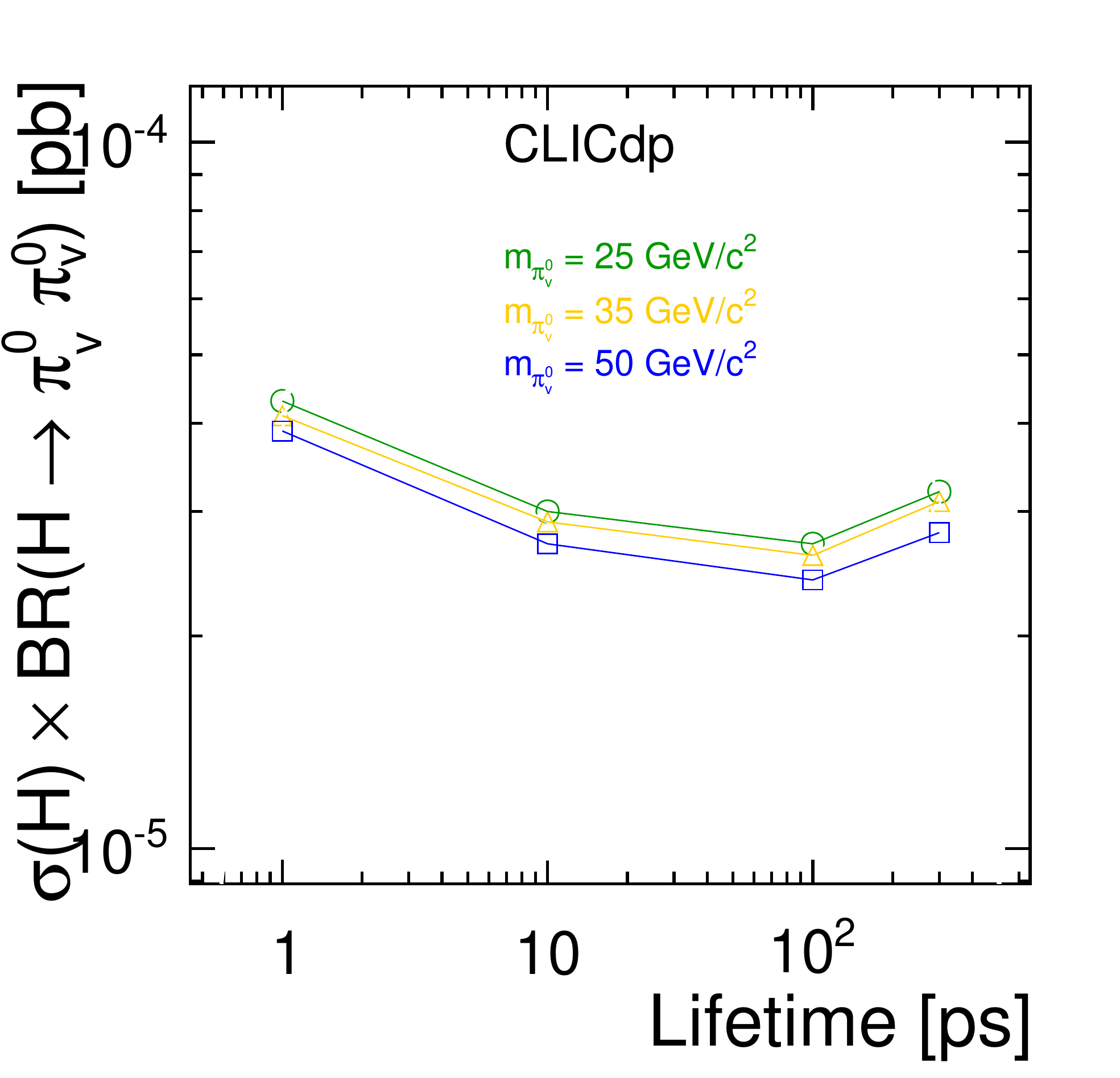}

}~~\subfloat[\label{fig:WIMPbg}LHC (blue) and CLIC (orange) exclusions for the
particles responsible of the generation of the baryon number in the
WIMP baryogenesis scenario. CLIC outermost (innermost) contour is
for 30 events (3 events) produced in the detector acceptance for displaced
vertexes.]{\begin{centering}
\includegraphics[width=0.49\linewidth]{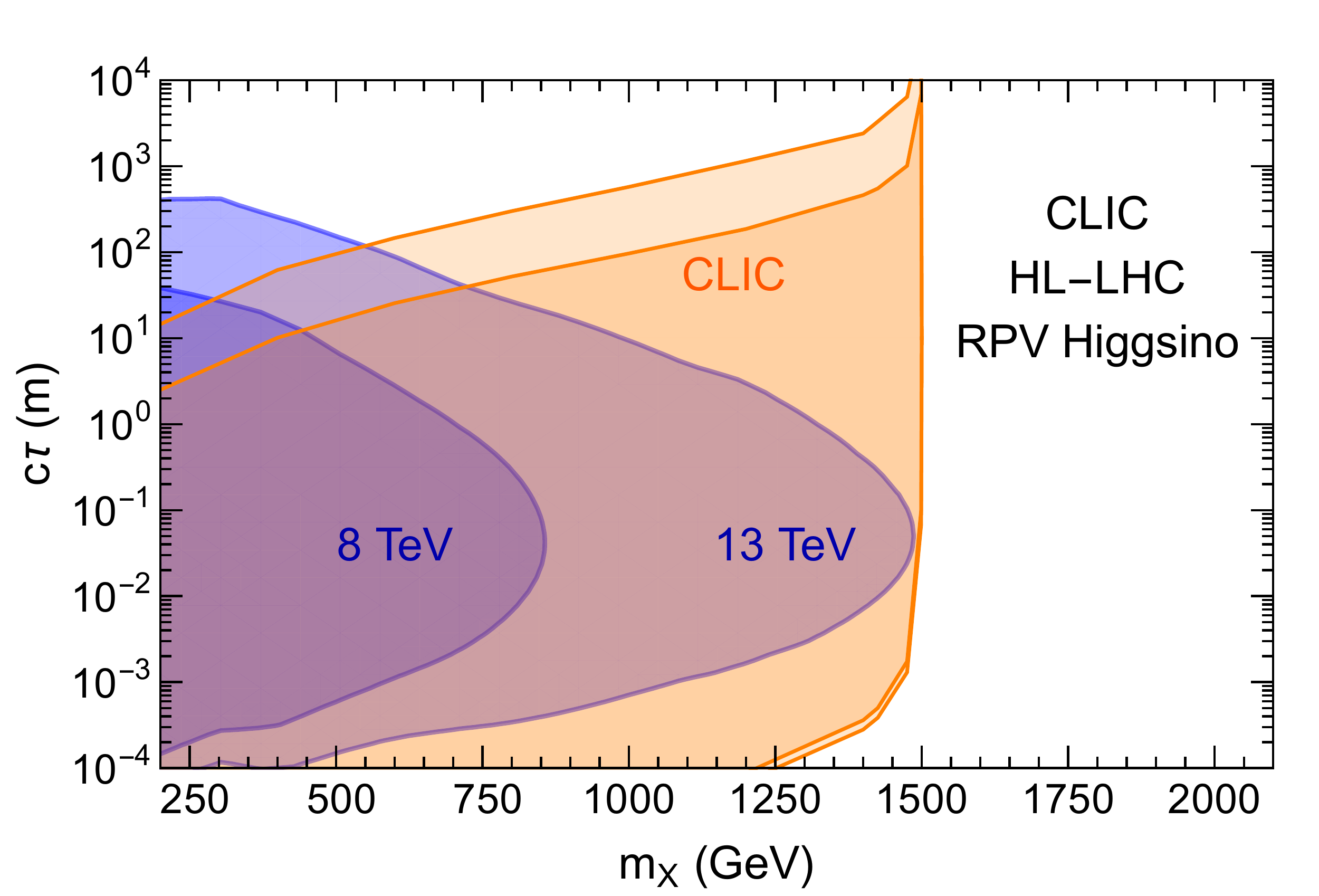}
\par\end{centering}
}
\par\end{centering}
\caption{Searches for Long Lived new physics states.}
\end{figure}

CLIC can also probe TeV-scale Baryogenesis models of radically different
nature. In particular it is possible to test the ``WIMP baryogenesis''
scenario \cite{Cui:2012ve}, where the baryon asymmetry is generated
via the baryon number violating decays of TeV-scale long-lived particles.
The favorable experimental conditions of CLIC allow to probe unexplored
regions of the mass-lifetime parameter space of this model \cite{1812.02093}.
As shown in Figure~\ref{fig:WIMPbg} CLIC can explore long decay
lengths that are necessary to generate necessary out-of-equilibrium
decays in the early Universe, significantly extending the reach of
the HL-LHC~\cite{Cui:2013aa,Cui:2015aa,Cui:2012ve,Cui:2015qy}.

\paragraph{Hidden Sector}

The possibility that sets of particles secluded from our view exist
in ``mirror world'' is an open question. These particles may be
secluded to us because of a tiny coupling between Standard Model states
and the new physics states in question. Such feeble interactions may
be useful in a number of contexts to address open issues of the Standard
Model, see e.g. Ref.~\cite{Alexander:2016lr} for a discussion, hence
their search is very motivated. These searches are very challenging
because the properties of the new physics states can only be vaguely
guessed, hence a broad program of searches needs to put in place to
effectively explore this idea. In this context it is possible that
new physics manifests itself with light new particles, which we have
not yet seen because of their tiny couplings with SM particles. CLIC
can make progress on the experimental exploration of this scenario
in unique corners of its vast parameter space. For example, the clean
environment and the absence of trigger allows CLIC to improve significantly
over the HL-LHC in the search for Higgs or Higgs-like bosons decay
to long-lived particles \cite{Kucharczyk:2018aa,Kucharczyk:2625054},
reaching exclusion of 
\[
BR(h\to\text{displaced vertexes})\sim10^{-4},
\]
as shown in Figure~\ref{fig:Excluded-cross-section-h2PiV}. This
result can be recast for searches of heavier Higgs bosons, as detailed
in Section 8 of Ref.~\cite{1812.02093}. These searches allow CLIC
to probe models of electroweak symmetry breaking, emerged in the context
of ``Neutral Naturalness'' scenario, such as the Fraternal Twin
Higgs \cite{Craig:2015qv} and Folded Supersymmetry (see e.g \cite{Burdman:2007aa})
solution of the Naturalness Problem.

\begin{figure}
\begin{centering}
\includegraphics[width=0.49\linewidth]{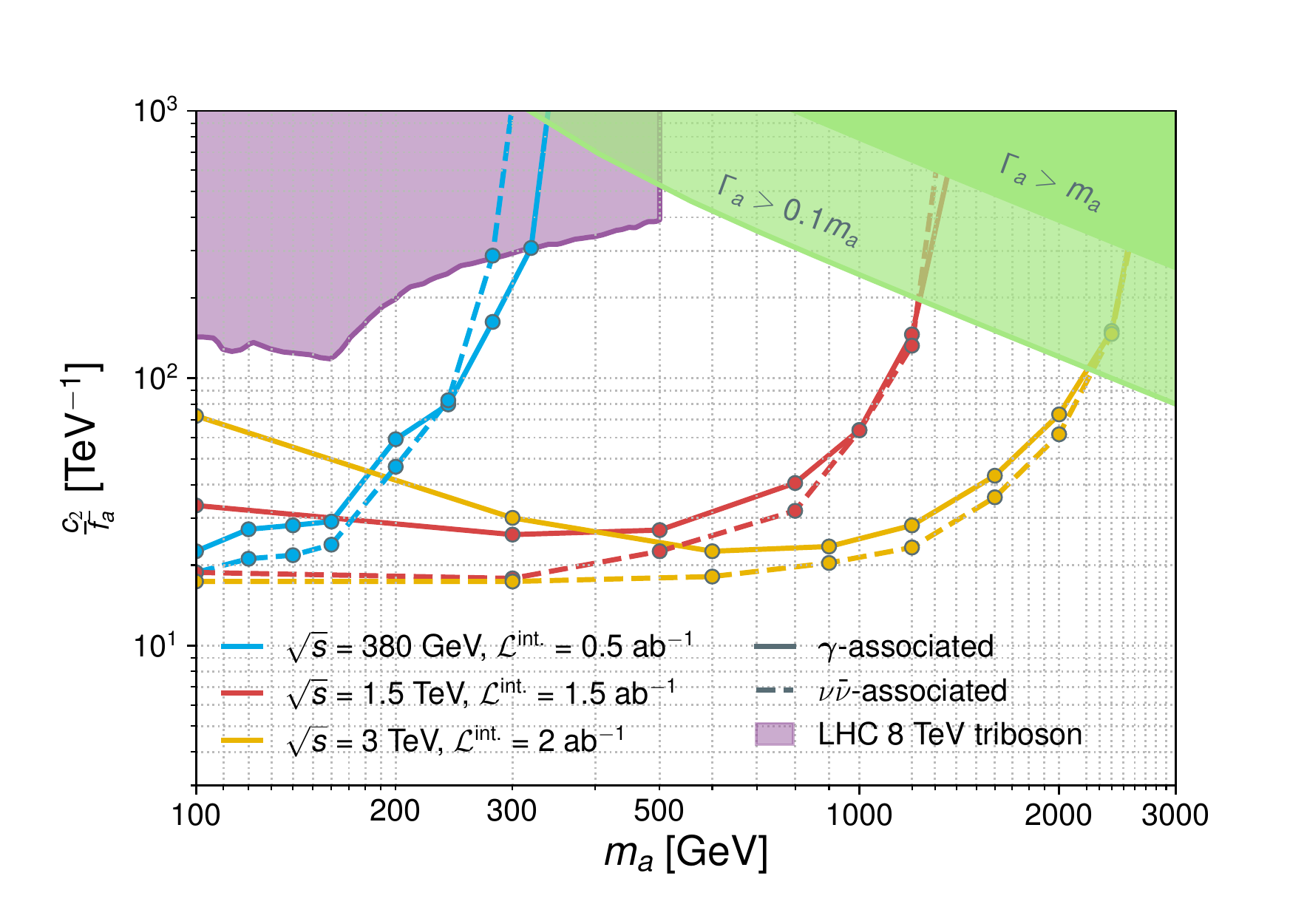}
\par\end{centering}
\caption{\label{fig:CLIC-ALP-photophobic-exclusion}CLIC 95\% exclusion reach
from Ref.~\cite{1812.02093} for a heavy axion-like particle in the
photo-phobic limit.}
\end{figure}

CLIC can also search for relatively heavy Axion-Like Particles, that
may be part of a feebly interacting sector that extends the Standard
Model. These sectors may find their origin in several theoretical
contexts, hence they are a useful simplified model to express the
reach of CLIC in general parameter space ruled by the mass of the
ALP and its decay constant. As a high energy collider CLIC can probe
ALPs that are obviously outside the reach of dedicated low-energy
experiments~\cite{Bauer:2018ab}. In Figure~\ref{fig:CLIC-ALP-photophobic-exclusion}
we show results for the photo-phobic ALP \cite{Craig:2018aa} case,
that are in any case representative for other couplings structures
involving photons as well, and we can see about one order of magnitude
of improvement in the bound of the decay constant of the ALP. In particular
it is remarkable that CLIC can improve on LHC bounds and is able to
enter regions of parameters space for the model in which the ALP mass
is less than or comparable with its decay constant, where the models
are more motivated.

Last but not least it should be recalled that CLIC technology may
already now enable new types of experiments with low energy electron
beams that may discover new light gauge bosons of possible relevance
for the Dark Matter puzzle~\cite{Akesson:2018ab,Akesson:2018aa}.
This might be a useful testing ground for CLIC technology with a very
good chance to add new knowledge in the quest for Dark Matter. On
a similar note it is worth to recall that a multi-TeV leptonic collider
may allow parasitic use of the beam for fixed target experiments~\cite{Kanemura:2015kq}
sensitive to other types of light new particles, therefore attaching
a bonus exploration of the so-called ``intensity frontier'' to the
already very rich exploration of the high energy frontier that CLIC
can deliver.


\paragraph{Neutrino Mass}

\begin{figure}
\begin{centering}
\includegraphics[width=0.49\linewidth]{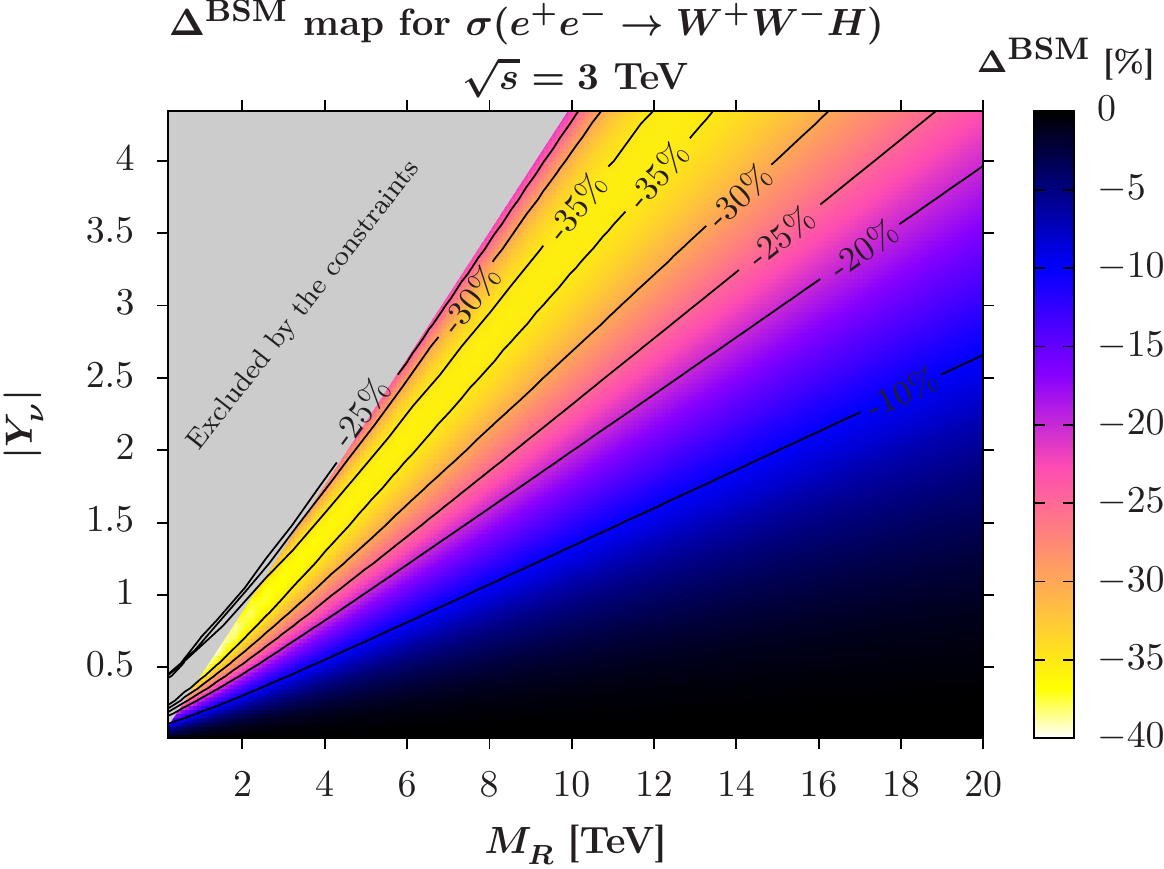}
\par\end{centering}
\caption{\label{fig:Relative-change-WWH} Relative change on the $WWH$ cross-section
due to a right-handed neutrino of mass $M_{R}$ and inverse-seesaw
Yukawa coupling $Y_{\nu}$ \cite{Baglio:2019aa,Baglio:2017ab}.}
\end{figure}

Evidence for flavor oscillations in neutrinos demands these particles
to be massive. Mixing parameters of the neutrino flavors subject to
weak interactions with the charged leptons are established in experiments.
Still we lack a deep understanding of the origin of neutrino masses.
These masses cannot be explained in the Standard Model and they require
new physics either in the form of a chiral partner of the left-handed
neutrinos, that is as a heavy right-handed neutrino, or in the form
of new contact interactions between leptons and the Higgs boson. These
interactions happen to be non-renormalizable, hence they require some
new physics at higher mass scales to originate to them.

CLIC has sensitivity to a large set of models in which lepton number
is an almost approximate symmetry, which makes very natural to have
small neutrino masses in the form of a Majorana mass. Other presentations
appeared at this workshop \cite{Bhupal-Dev:2019aa,Baglio:2019aa,Bhupal-Dev:2018aa,Baglio:2017ab}
give in-depth discussions on the models and the search for these models
at colliders. Here we recall some representative results. For example,
see Figure \ref{fig:Relative-change-WWH}, in the inverse-seesaw model
it is possible to have large Yukawa couplings between the Higgs boson
and the right-handed neutrinos, which CLIC can exclude up to 10 TeV
mass when the Yukawa coupling is of order 1 \cite{Baglio:2019aa,Baglio:2017ab}.
Even greater reach around tens of TeV is expected for models (e.g.
these discussed in Refs. \cite{King:2014aa,Geib:2016aa,Babu:2009aa})
featuring a doubly charged scalar lepton for Yukawa coupling of order
1 \cite{Crivellin:2018aa}. Furthermore CLIC can easily exclude the
presence of electroweak charged scalars and fermions, such as the
heavy mediators of type-2 and type-3 see-saw models, with masses below
1.5 TeV over the entire parameters space of the models. In particular
for type-2 sees-saw CLIC is able to probe the model for any value
of the vacuum expectation of the triplet neutral scalar \cite{Agrawal:2018aa},
dramatically improving on the situation of the HL-LHC which is hardly
sensitive to the case of VEV greater than 100~KeV \cite{Antusch:2018aa}.

\paragraph{}

\section{Summary of physics results, their impact and discussion}

The above results make clear that a multi-TeV lepton collider emerges
as a uniquely powerful and balanced option for future exploration
of the high energy frontier. Multi-TeV lepton colliders can attain
a thorough exploration of the TeV scale and deliver significant progress
on the several open issues of the Standard Model. In fact, CLIC can
throughly explore the existence of new Higgs bosons in the TeV mass
ballpark observing them directly as new particles produced in the
$\ee$ collisions as well as measuring their subtle impact on the
couplings of the 125 GeV Higgs boson. Furthermore CLIC can test TeV
scale solutions of the baryon number and Dark Matter puzzles which
involve TeV scale new states as well as test neutrino mass generation
mechanisms featuring new dynamics at the TeV scale.

The strength of CLIC reach for new physics is built on two pillars:
the possibility to carry out \emph{very precise measurements}, that
is typical of $\ee$ machines, and the unprecedented large center
of mass energy it can attain, 3 TeV, which makes it a \emph{discovery
machine} capable to observe a large set of the new particles that
are predicted in motivated new physics models. In the discovery of
new particles CLIC is remarkable because of its ability to explore
directly the electroweak sector up to multi-TeV energy scales. This
is a unique and defining feature of CLIC in the landscape of currently
proposed $\ee$ projects.

The interpretation of the new physics reach of CLIC in the context
of new physics scenarios makes clear that CLIC competes very well
with the reach of $pp$ colliders. In fact the possibility to search
for new electroweak states enables to test new physics scenarios searching
for their very core ingredients, i.e. CLIC is often sensitive to the
electroweak states that are directly involved in the solutions to
the open issues of the Standard Model. The interpretation of searches
for colored particles, at any collider, is in general less sharp and
less conclusive, as these states can relatively easily be avoided
or made heavier in most models. Therefore it appears fair to say that
a multi-TeV $\ee$ machine such as CLIC can teach us sharp and definitive
lessons on the physics of the TeV.

The great advantage of CLIC in the search for new electroweak states
comes from the relatively low levels of backgrounds that are expected.
In such an environment it will be possible to test thoroughly new
physics at the weak scale, including exotic kinds of new physics that
show up in subtle signatures such as anomalous tracks, extra displaced
vertexes in the events, other types of long lived states or exotic
jet-like energy deposits recorded only in some of the layers of the
detectors. Being sensitive to new electroweak charged states in the
whole range of energies up to beyond TeV masses, including when they
give rise to subtle and exotic signatures, CLIC will bring us to a
vantage point from which we can draw definitive conclusions on many
open issues of the Standard Model.

It should also be remarked that CLIC will deliver th\emph{e full set
of results in a relatively short time-scale}, only 34 years from the
first stone to the last recorded collision, which is a relatively
short time in the landscape of the future colliders projects currently
under discussion. This gives best chances to keep an active experimental
community, constantly reinvigorated by young ones attracted by the
challenge of working on intellectually stimulating projects at the
technical forefront of the epoch. Furthermore it is possible to envisage
a fruitful concurrence with other lines of experiments that attack
the quest for fundamental interactions from a different angle, e.g.
small and large dark matter detectors or highly-precise low energy
experiment, which may otherwise preempt the scope of discovery of
a future collider facility if this is coming on-line or delivering
final results too late. On the flip side, the interplay with other
experimental lines of research may be very synergetic as the CLIC
energy and luminosity plans can be adjusted to follow hints from other
searches for new physics.


\providecommand{\href}[2]{#2}\begingroup\raggedright\endgroup

\end{document}